\documentclass[12pt]{article}
\usepackage{amsfonts,amssymb,amscd}
\usepackage{amsmath}
\usepackage[subnum]{cases}
\usepackage{float}
\usepackage{color}
\usepackage{cite}

\def\1o2{{1\over2}}

\usepackage{graphicx}

\begin{document}

\title{\textbf{\\ Hairy Kiselev Black Hole Solutions}}
\author{ Yaghoub Heydarzade$^{(a)}$\footnote{yheydarzade@bilkent.edu
.tr}, Maxim Misyura $^{(b,c)}$\footnote{max.misyura94@gmail.com}\,\,
and Vitalii Vertogradov$^{(c,d)}$\footnote{vdvertogradov@gmail.com} \\
\\
{\small (a) Department of Mathematics, Faculty of Sciences, Bilkent University, 06800 Ankara, Turkey}\\
{\small (b) Department of High Energy and Elementary Particles Physics,
  Saint Petersburg State University,}\\
  {\small University Emb. 7/9, Saint Petersburg, 199034, Russia}\\
{\small (c) Physics department, Herzen state Pedagogical University of Russia,
48 Moika Emb.,} \\ {\small Saint Petersburg 191186, Russia} \\
{\small (d) SPB branch of SAO RAS, 65 Pulkovskoe Rd, Saint Petersburg
196140, Russia}
}

%\date{\today}
%\date{March 20, 2016}

\begin{titlepage}
\maketitle
\thispagestyle{empty}

\begin{abstract}

In the realm of astrophysics, black holes exist within non-vacuum cosmological backgrounds, making it crucial to investigate how these backgrounds influence the properties of black holes. In this work,
we first introduce a novel  static spherically-symmetric exact solution of Einstein field equations representing a surrounded  hairy black hole. This solution represents a generalization of the hairy Schwarzschild solution recently derived using the extended gravitational decoupling method. Then,
we discuss how the new induced modification terms attributed to the primary hairs and various background fields  affect the geodesic motion in comparison
to the  conventional Schwarzschild case. Although these
modifications may appear insignificant in most cases, we identify specific conditions where they can be
comparable to the Schwarzschild case for some particular background fields.
\\


\end{abstract}

\end{titlepage}
%\pacs{04.20.Jb, 04.40.Nr}

\setcounter{page}{2}
\section{Introduction}

In 2019 the Event Horizon Telescope Collaboration unveiled the very first image of a black hole located at the center of the massive elliptical galaxy  M87 \cite{bib:9, bib:10, bib:11}. More recently, scientists have successfully observed the shadow of the supermassive black hole located in the center of our own galaxy \cite{bib:ehtc2022}. These direct observations provide compelling evidence that black holes are not merely abstract mathematical solutions of the Einstein field
equations but real astrophysical objects. Black holes possess a range of
miraculous properties. For instances, they allow for the extraction of
energy from their rotation and electric fields \cite{bib:pen, bib:zaslav, bib:mp, bib:charged_vaidya}.
In the vicinity of the black hole's event horizon, particles can possess negative energy \cite{bib:pen, bib:zaslav_rn, bib:mp, bib:grib, bib:ver_negative,
    bib:ver_ker_negative}, and  black holes can even function  as particle
accelerators~\cite{bib:bsw, bib:zaslav_anti, bib:zaslav_dirty,
    bib:grib_complex, bib:ver_complex, bib:joshi_col, bib:japan_col}. 
\\
\\
In the realm of astrophysics, black holes are not isolated objects and
they inhabit non-vacuum backgrounds. Some research has focused on
investigating the direct local effects of cosmic backgrounds on the
known black hole solutions. For instance, Babichev et al. \cite{babi}
have shown that in an expanding universe by a phantom scalar field, the
mass of a black hole decreases as a result of the accretion of particles
of the phantom field into the central black hole. However, one notes
that this is a global impact. To explore the local changes in the
spacetime geometry near the central black hole, one should consider a
modified metric that incorporates the surrounding spacetime. In this
context, an analytical static spherically symmetric solution to Einstein
field equations has been presented by Kiselev \cite{bib:50}. This
solution generalizes the
usual Schwarzschild black hole to a nonvacuum background and is characterized by an effective equation of state parameter of the surrounding field of the black hole. Hence it can encompass a wide range of possibilities including quintessence, cosmological constant, radiation and  dust-like fields. Several properties of the Kiselev black hole have been extensively investigated in the literature [85-90].
 Later, this solution has been generalized to the dynamical  Vaidya type
solutions~\cite{bib:kvaidya, bib:kbvaidya, bib:kevaidya}.
Such generalizations are well justified due to the non-isolated nature of real-world black holes and their exitance in non-vacuum backgrounds. Black hole solutions coupled to matter fields, such as the Kiselev solution, are particularly relevant for the study of astrophysical black holes with distortions ~\cite{bib:51, bib:52, bib:53,
bib:54}. They also play a significant role in investigating the no-hair theorem \cite{bib:55, bib:56, bib:57, bib:58}. This theorem states that a black hole can be described only with three charges (i.e.
 mass $M$, electric charge $Q$ and angular momentum $a$), and it relies
on a crucial assumption that the black hole is isolated, meaning that
the spacetime is asymptotically flat and free from other sources.
However, real-world astrophysical situations do not meet this assumption
. For instances, one may refer to black holes in binary systems, black
holes surrounded by plasma, or those accompanied by accretion disks or
jets in their vicinity. Such situations imply that a black hole may
put on different types of wigs, and hence the applicability of the
standard no-hair theorem for isolated black holes to these cases becomes
questionable \cite{bib:56, bib:57, bib:58, bib:noh4, bib:hok_hair}.

Recently,  the minimal geometrical deformations~\cite{bib:mgd1,bib:mgd2, bib:mgd3} and the extended gravitational decoupling
methods~\cite{bib:gd1, bib:gd2, bib:gd3}  have been utilized to derive new
solutions from the known seed solutions of Einstein field equations.
These techniques have been particularly effective in investigating the
violation of the no-hair theorem, the emergence of novel types of hairy
black holes, and the exploration of alternative theories of
gravity~\cite{bib:r1, bib:r2, bib:r3, bib:r4, bib:r5, bib:r6, bib:r7,
    bib:r8}
Using the extended gravitational
decoupling method, Ovalle et all~\cite{bib:bh1}
have introduced a generalization of a Schwarzschild black hole
surrounded by an anisotropic fluid and possesses primary hairs.
This new solution has motivated  a substantial further research in generalizing this solution to hairy
Kerr~\cite{bib:hairy_kerr}, Vaidya and generalized
Vaidya~\cite{bib:vermax}, regular hairy  black holes~\cite{bib:vermax2,
    bib:ovalle_regular} and many others.
Indeed, the gravitational decoupling
method represents a novel and powerful tool for obtaining new solutions to the
Einstein equations.\\
\\
In the present work, we introduce a novel class of exact solutions to the Einstein field equations, which describe a surrounded
hairy  Schwarzschild black hole. This solution serves as a generalization of the previously obtained hairy Schwarzschild solution using the extended gravitational decoupling method.  Then, in order to analyze the properties
of the solution, we investigate the effect of the new modification terms, attributed to the primary hairs and various surrounding fields, on the timelike
geodesic motion. Specifically, we compare the effects of modification
terms to the conventional Schwarzschild case. While these modifications
may seem negligible in most scenarios, we identify specific situations
where they can be comparable to the Schwarzschild case, particularly
when specific surrounding fields are present. This analysis sheds light
on the significance of these modifications in certain situations,
providing insights into the behavior of geodesic motion around real
astrophysical black holes.
\\

The structure of the present paper is as follows. In Section 2, we briefly discuss the
hairy Schwarzschild solution by the minimal geometrical deformations and
the extended gravitational decoupling method. In Section 3,  we solve the Einstein
field equations in order to obtain the surrounded hairy Schwarzschild black hole. In Section 4, we do analysis of the timelike geodesic
motion. In  Section 5, we summarize the new findings and implications of the study.
The system of units $c=G=1$ will be used throughout the paper.

\section{Gravitational decoupling and hairy Schwarzschild black hole}

Gravitational decoupling method states that one can solve the Einstein
field equations with the matter source
\begin{equation}
\tilde{T}_{ik}=T_{ik}+\Theta_{ik} \,,
\end{equation}
where $T_{ik}$ represents the energy-momentum tensor of a system for
    which the Einstein field equations are
\begin{equation} \label{eq:thefirst}
G_{ik}=8\pi T_{ik} \,.
\end{equation}
The solution of the equations \eqref{eq:thefirst} is supposed to be
known and represents the seed solution.
Then $\Theta_{ik}$ represents an extra matter sources which causes
additional geometrical deformations.
The Einstein equations for this
new matter source are
\begin{equation} \label{eq:thesecond}
\bar{G}_{ik}=\alpha \Theta_{ik} \,,
\end{equation}
where $\alpha$ is a coupling constant and $\bar{G}_{ik}$ is the Einstein
tensor of deformed metric only.
 The gravitational decoupling method
states that despite of nonlinear nature of the Einstein equations, a
straightforward superposition  of these two solutions
\eqref{eq:thefirst} and \eqref{eq:thesecond}
\begin{equation}
\tilde{G}_{ik}\equiv G_{ik}+\bar{G}_{ik}=8\pi T_{ik}+\alpha
\Theta_{ik}\equiv \tilde{T}_{ik} \,,
\end{equation}
is also the solution of the Einstein field equations.

Now, we briefly describe this method. Let us consider the Einstein field
equations,
\begin{equation} \label{eq:ex1}
G_{ik}=R_{ik}-\frac{1}{2}g_{ik}R=8\pi T_{ik} \,.
\end{equation}
Let the solution of \eqref{eq:ex1} is a static spherically-symmetric
spacetime of the form
\begin{equation} \label{eq:seed}
ds^2=-e^{\nu(r)}dt^2+e^{\lambda(r)}dr^2+r^2 d\Omega^2 \,.
\end{equation}
Here $d\Omega^2=d\theta^2+\sin^2\theta d\varphi^2$ is the metric on unit
two-sphere, $\nu(r)$ and $\lambda(r)$ are functions of $r$ coordinate
only, and they are supposed to be known.
The metric \eqref{eq:seed} is
termed as the seed metric.

Now, we seek the geometrical deformation of \eqref{eq:seed} by
introducing two new functions $\xi=\xi(r)$ and $\eta=\eta(r)$ by:
\begin{eqnarray}
 \label{eq:deform}
&&e^{\nu(r)} \rightarrow e^{\nu(r)+\alpha \xi(r)}, \nonumber\\
&&e^{\lambda(r)} \rightarrow e^{\lambda(r)}+\alpha \eta(r).
\end{eqnarray}
Here $\alpha$ is a coupling constant. Functions $\xi$ and $\eta$ are
associated with the geometrical deformations of $g_{00}$ and $g_{11}$ of the
metric \eqref{eq:seed}, respectively. These deformations are caused by
new matter source $\Theta_{ik}$. If one puts $\xi(r)\equiv 0$ then the
only $g_{11}$ component is deformed, leaving $g_{00}$ unperturbed -- this
is known as the minimal geometrical deformation. It has some drawbacks, for
example, if one considers the existence of a stable black hole possessing
a well-defined event horizon~\cite{bib:mgd3}. Deforming both
$g_{00}$ and $g_{11}$ components  is an arena of the extended
gravitational decoupling. One should note that gravitational decoupling
can lead to an energy exchange between two matter
sources~\cite{bib:ovalle_new}. For example, if one opts for
gravitational decoupling of Vaidya spacetime, then one can decouple the
usual Vaidya spacetime without energy exchange. However, in the generalized
Vaidya spacetime, there is an energy exchange for arbitrary mass function
$M
(v,r)$~\cite{bib:vermax}.

Substituting \eqref{eq:deform} into \eqref{eq:seed}, one obtains
\begin{equation} \label{eq:noseed}
ds^2=-e^{\nu+\alpha \xi}dt^2+\left(e^{\lambda}+\alpha \eta \right)
dr^2+r^2 d\Omega^2 \,.
\end{equation}
The Einstein equations for \eqref{eq:noseed} as\begin{equation}
\tilde{G}_{ik}= 8 \pi \tilde{T}_{ik}=8\pi (T_{ik}+\Theta_{ik} ) \,,
\end{equation}
give
\begin{eqnarray} \label{eq:einstein}
&&8\pi (T^0_0+\Theta^0_0)=-\frac{1}{r^2}+e^{-\beta}\left( \frac{1}{r^2}-\frac{\beta'}{r}\right) ,\nonumber \\
&&8\pi (T^1_1+\Theta^1_1)=-\frac{1}{r^2}+e^{-\beta}\left(\frac{1}{r^2}+\frac{\nu'+\alpha \xi'}{r}\right), \nonumber \\
&&8\pi (T^2_2+\Theta^2_2)=\frac{1}{4}e^{-\beta}\left(2(\nu''+\alpha \xi'')+(\nu'+\alpha \xi')^2-\beta'(\nu'+\alpha \xi')+2\frac{\nu'+\alpha \xi'-\beta'}{r} \right), \nonumber\\
&&e^{\beta}\equiv e^{\lambda}+\alpha \eta.
\end{eqnarray}
Here the prime sign denotes the partial derivative with respect to the radial
coordinate $r$, and we have $8\pi\left( T^2_2+\Theta^2_2\right)=8 \pi \left(T^3_3+\Theta^3_3\right)$ due to
the spherical symmetry.

From \eqref{eq:einstein} one can define the effective energy density
$\tilde{\rho}$, effective radial and tangential
 $\tilde{P}_r$, $\tilde{P}_t$ pressures as
\begin{eqnarray} \label{eq:effective}
&&\tilde{\rho}=-(T^0_0+\Theta^0_0), \nonumber \\
&&\tilde{P}_r=T^1_1+\Theta^1_1, \nonumber \\
&&\tilde{P}_t=T^2_2+\Theta^2_2. 
\end{eqnarray}
From \eqref{eq:effective} one can introduce the anisotropy parameter
$\Pi$ as
\begin{equation} \label{eq:anisotropy}
\Pi=\tilde{P}_t-\tilde{P}_r \,,
\end{equation}
where if $\Pi\neq 0$ then it indicates the anisotropic behaviour of fluid
$\tilde{T}_{ik}$.

The equations \eqref{eq:einstein} can be decoupled into two
parts\footnote{One should remember that it always works for
    $T_{ik}\equiv 0$ i.e. the vacuum solution and for special cases of
    $T_{ik}$ if one opts for Bianchi identities
    $\nabla_iT^{ik}=\nabla_i\Theta^{ik}=0$ with respect to the metric
    \eqref{eq:noseed} otherwice there is an energy exchange i.e.
    $\nabla_i\tilde{T}^{ik}=0\Rightarrow
    \nabla_iT^{ik}=-\nabla_i\Theta^{ik}\neq 0$.}:
the Einstein equations corresponding to the seed solution \eqref{eq:seed} and the
one corresponding to the geometrical deformations.
If we consider the
vacuum solution i.e. $T_{ik}\equiv 0$ - Schwarzschild solution, then by
solving the Einstein field equations which correspond the geometrical
deformations, one obtains the hairy Schwarzschild  solution~\cite{bib:bh1}
\begin{equation} \label{eq:seedsch}
ds^2=-\left(1-\frac{2M}{r}+\alpha e^{-\frac{r}{M-\frac{\alpha
l}{2}}}\right) dt^2+\left(1-\frac{2M}{r}+\alpha e^{-\frac{r}{M-\frac{\alpha
l}{2}}}\right) ^{-1}dr^2+r^2d\Omega^2 \,,
\end{equation}
where $\alpha$ is the coupling constant, $l$ is a new parameter with
length dimension and associated with a primary hair of a black hole. Here
$M$
is the mass of the black hole in relation with the
Schwarzschild mass $\mathcal{M}$ as
\begin{equation}
M=\mathcal{M}+\frac{\alpha l}{2} \,.
\end{equation}
The impact
of $\alpha$ and $l$ on the geodesic motion, gravitational lensing,
energy extraction   and the thermodynamics has been studied
in Refs.\cite{bib:geod, bib:lens, bib:energy, bib:thermo, 
    bib:ver_thermo}, and the influence of primary hair on quasinormal
frequencies for scalar, vector and tensor perturbation fields has been
investigated in \cite{bib:contreras}.

\section{Surrounded hairy Schwarzschild black hole}

Recently, the hairy Schwarzschild black hole has been
introduced in~\cite{bib:bh1} by using the gravitational decoupling
method.
This solution in the Eddington-Finkelstein coordinates  takes the
form
\begin{equation} \label{eq:hairysch}
ds^2=-\left(1-\frac{2M}{r}+\alpha e^{-\frac{r}{M-\frac{\alpha
l}{2}}}\right) dv^2+2\varepsilon dvdr+r^2d\Omega^2 \,.
\end{equation}
Here $v$  is the advanced $(\varepsilon=+1)$ or retarded $(\varepsilon=-1)$
Eddington time. In this section, using the approach in~\cite{bib:50, bib:106,
bib:kvaidya}, we obtain the generalization of  this solution
 representing a hairy Schwarzschild solution
surrounded by some particular fields motivated by cosmology as in the following
theorem. 

\textbf{Theorem:} \textit{Considering the extended gravitational
decoupling \cite{bib:gd2} and the principle of additivity and linearity
in the energy-momentum
tensor \cite{bib:50} which allows one to get correct limits to the known solutions, the
Einstein field equations admit the following solution in the Eddington-Finkelstein coordinates
\begin{equation} \label{eq:methair*}
ds^2=-\left(1-\frac{2M}{r}-\frac{N}{r^{3\omega+1}}+\alpha
e^{-\frac{2r}{2M-\alpha l}} \right) dv^2+2\varepsilon dvdr +r^2d\Omega^2
\,,
\end{equation}
where $M=\mathcal{M}+\frac{\alpha l}{2}$ in which $\mathcal{M}$ and $\mathcal{M}$ are integration constants. The
\textit{metric represents a surrounded hairy Schwarzschild  solution} or
    equivalently hairy Kiselev solution.} We summarize our proof as follows.

Let us consider the general spherical-symmetric spacetime in the form
\begin{equation} \label{eq:metgen}
ds^2=-f(r)dv^2+2\varepsilon dvdr+r^2d\Omega^2 \,.
\end{equation}
The Einstein tensor components for the metric \eqref{eq:metgen} are
given by
\begin{eqnarray} \label{eq:genei}
&&G^0_0=G^1_1=\frac{1}{r^2}\left(f'r-1+f\right) \,, \nonumber\\
&&G^2_2=G^3_3=\frac{1}{r^2}\left(rf'+\frac{1}{2}r^2f''\right)  \,,
\end{eqnarray}
where the prime sign represents the derivative with respect to the radial
coordinate $r$.
The total energy-momentum tensor should be a combination  of
$\Theta_{ik}$ associated to the minimal
geometrical deformations and $T_{ik}$  associated to the surrounding
 fluid as
\begin{equation} \label{eq:temtotal}
\tilde{T}_{ik}=\alpha \Theta_{ik}+T_{ik} \,.
\end{equation}
One should note that here we do not demand the fulfilment of the
condition $\Theta^{ik}_{;k}=T^{ik}_{;k}=0$.
Instead, we demand that
$\tilde{T}^{ik}_{;k}=0$ which follows the Bianchi identity.
The total energy-momentum tensor $\tilde{T}_{ik}$ follows the same
symmetries of the Einstein tensor  (\ref{eq:genei}) for
(\ref{eq:metgen}), i.e.,\,
$\tilde{T}^0_0=\tilde{T}^1_1$ and $\tilde{T}^2_2=\tilde{T}^3_3$.

An appropriate general expression for the energy-momentum tensor $T_{ik}$ of the surrounding
fluid can be~\cite{bib:50}
\begin{eqnarray} \label{eq:temsur}
&&T^0_0=-\rho(r) \,,\nonumber\\ 
&&T^i_k= -\rho(r)\left[ -\xi \left(1+3\zeta \right)
\frac{r^ir_k}{r^nr_n}+\zeta \delta^i_k\right] \,.
\end{eqnarray}
From the form of the energy-momentum tensor \eqref{eq:temsur}, one can
see that the spatial profile is proportional to the time component,
describing the energy density $\rho$ with arbitrary constants $\xi$ and
$\zeta$ depending on the internal structure of the surrounding fields.
The isotropic averaging over the angles results in
\begin{equation}
<T^i_k>=\frac{\xi}{3}\rho \delta^i_k=P\delta^i_k \, ,
\end{equation}
since we considered $<r^{i}r_{k}>=\frac{1}{3}{\delta^{i}}_{k}r_n r^n$.
Then, we have a barotropic equation of state for the surrounding fluid
as
\begin{equation} \label{eq:alphadef}
P(r)=\omega \rho(r)~, \,~~ \omega=\frac{\xi}{3} \,,
\end{equation}
where $P(r)$ and $\omega$ are the pressure and the constant equation of the
state parameter of the surrounding field, respectively.
Here, one notes that the source  $T_{ik}$  associated to the surrounding
fluid should possess the same symmetries in $\tilde{T}_{ik}$  because
$\Theta_{ij}$ associated to the
geometrical deformations has the same symmetries as \footnote{One should
note that hairy Schwarzschild solution is supported with an anisotropic
fluid $\Theta^i_k$
\begin{equation}
\Theta^0_0=-\bar{\rho} \,, \Theta^1_1=\bar{P}_r\,, \Theta^2_2=\Theta^3_3=\bar{P}_t \,.
\end{equation}
Where the non-vanishing parameter $\Pi=\bar{P}_t-\bar{P}_r$ indicates
    on the anisotropic nature of the energy momentum tensor. So, in order to
    satisfy the condition $\Theta^0_0=\Theta^1_1$ the anisotropic fluid
    should be satisfied with the equation of the state $P_r=-\bar{\rho}$.}
\begin{eqnarray}\label{thetacomp}
&&\Theta^0_0=\Theta^1_1=-\bar{\rho},\nonumber\\
&&\Theta^2_2=\Theta^3_3=\bar{P}_{t}.
\end{eqnarray}
It means that 
${T}^0_0={T}^1_1$ and ${T}^2_2={T}^3_3$.
These
exactly provide the so-called principle of additivity and linearity
considered in~\cite{bib:50} in order to determine the free
parameter $\zeta$ of the energy-momentum tensor $T_{ik}$ of surrounding
fluid as
\begin{equation} \label{eq:zetadef}
\zeta=-\frac{1+3\omega}{6\omega} \,.
\end{equation}
Now, substituting \eqref{eq:alphadef} and \eqref{eq:zetadef} into
\eqref{eq:temsur}, the nonvanishing components of the surrounding
energy-momentum tensor $T_{ik}$ become
\begin{eqnarray}\label{tcomp}
&&T^0_0=T^1_1=-\rho, \nonumber \\
&&T^2_2=T^3_3=\frac{1}{2}\left(1+3\omega \right) \rho \,.
\end{eqnarray}
Now, we know the Einstein tensor components \eqref{eq:genei} and the
total energy-momentum tensor 
\eqref{eq:temtotal}. Putting all these equations together, the
$G^0_0=\tilde{T}^0_0$ and $G^1_1=\tilde{T}^1_1$ give us the following
equation:
\begin{equation} \label{eq:firstdif}
\frac{1}{r^2}\left( f'r-1+f\right)=-\rho-\alpha\bar{\rho} \,.
\end{equation}
Similarly,  the $G^2_2=\tilde{T}^2_2$ and $G^3_3=\tilde{T}^3_3$
components yields
\begin{equation} \label{eq:seconddif}
\frac{1}{r^2}\left(rf'+\frac{1}{2}f''r^2\right)=\frac{1}{2}\left
(1+3\omega \right) \rho+\bar{P} \,.
\end{equation}
Thus, there are four unknown functions $f(r), \,\rho(r), \,\bar{\rho}(r)$  and $\bar{P}$ that can be determined analytically
by the differential equations \eqref{eq:firstdif} and
\eqref{eq:seconddif} with the following ansatz:
\begin{equation} \label{eq:additionalcon}
f(r)=g(r)-\frac{\alpha l}{r}+\alpha e^{-\frac{2r}{2M-\alpha l}} \,.
\end{equation}
Then, by substituting \eqref{eq:additionalcon} into \eqref{eq:firstdif}
and \eqref{eq:seconddif} and using (\ref{thetacomp}) one obtains the following system of
linear differential equations \footnote{Here we apply the Einstein equation
        $\hat{G}^i_k=\alpha\Theta^i_k$ to eliminate $\tilde{\rho}$ and
        $\tilde{P}$. $\hat{G}^i_k$ is the Einstein tensor for the spacetime
\begin{equation}
ds^2=-\left(1-\frac{\alpha l}{r}+\alpha e^{-\frac{2r}{2M-\alpha l}}
\right)dv^2+2\varepsilon dvdr+r^2d\Omega^2 \nonumber.
\end{equation}} for unknowns $\rho(r)$ and $g(r)$
\begin{eqnarray} \label{eq:totaldif}
&&\frac{1}{r^2}\left( g'r-1+g\right)=-\rho,\nonumber \\
&&\frac{1}{r^2}\left(rg'+\frac{1}{2}g''r^2\right)=\frac{1}{2}\left
(1+3\omega \right) \rho\,.
\end{eqnarray}
This second order linear system can be integrated to give the metric function $g(r)$ as\begin{equation}
g(r)=1-\frac{2\mathcal{M}}{r}-\frac{N}{r^{3\omega+1}} \,,
\end{equation}
and the energy density $\rho(r)$ of the surrounding field as
\begin{equation} \label{eq:den}
\rho(r)=-\frac{3\omega N}{r^{3(\omega+1)}} \,.
\end{equation}
Here $\mathcal{M}$ and $N$ are constants of integration representing the
Schwarzschild mass and the surrounding field structure parameter,
respectively. By putting all these solutions together, we arrive at the
\textit{surrounded hairy Schwarzschild  solution} or equivalently \textit{hairy Kiselev solution} as 
\begin{equation} \label{eq:methair}
ds^2=-\left(1-\frac{2M}{r}-\frac{N}{r^{3\omega+1}}+\alpha
e^{-\frac{2r}{2M-\alpha l}} \right) dv^2+2\varepsilon dvdr +r^2d\Omega^2
\,,
\end{equation}
where $M=\mathcal{M}+\frac{\alpha l}{2}$.
From \eqref{eq:den}, one can see
that the weak energy condition
demands
that  parameters $\omega$ and $N$ have different signs.

\section{Timelike geodesics}

Considering the geodesic motion in spherically-symmetric spacetime, without loss of generality, one
can consider the equatorial plane
$\theta=\frac{\pi}{2}$.
The geodesic equations for the metric
\eqref{eq:metgen} can be obtained by varying the following action:
\begin{equation} \label{eq:action}
S= \int \mathcal{L} d\tau =\frac{1}{2}\int \left(-f\dot{v}^2
+2\varepsilon \dot{v}\dot{r}+r^2\dot{\varphi}^2 \right) d\tau \,,
\end{equation}
where the dot sign means the derivative with respect to the proper time
$\tau$.
The spacetime \eqref{eq:methair} is spherically-symmetric and hence, in
addition to the time-translation Killing vector $\frac{\partial}{\partial t
}$, there exists another Killing vector $\varphi^i=\frac{\partial}{\partial \varphi}$ and
the corresponding
conserved
quantity, the  angular momentum per mass, is given by
\begin{equation} \label{eq:angular}
\varphi^iu_i=\frac{\partial \mathcal{L}}{\partial
\dot{\varphi}}=r^2 \dot{\varphi} =L \,.
\end{equation}
Taking into account \eqref{eq:action} and \eqref{eq:angular}, one
obtains the following three geodesic equations
\begin{equation} \label{eq:angulargeo}
\dot{\varphi}=\frac{L}{r^2} \,,
\end{equation}
\begin{equation} \label{eq:timegeo}
-\frac{1}{2}f' \dot{v}^2+r\dot{\varphi}^2-\varepsilon \ddot{v}=0 \,,
\end{equation}
\begin{equation} \label{eq:radialgeo}
\varepsilon \ddot{r}=f\ddot{v}+f'\dot{v}\dot{r} \,,
\end{equation}
where the prime sign denotes the derivative with respect to the radial
coordinate $r$.
Substituting \eqref{eq:angulargeo} into
\eqref{eq:timegeo}, one obtains
\begin{equation}
f\ddot{v}=\frac{\varepsilon f L^2}{r^3}-\frac{1}{2}\varepsilon f f' \dot{v}^2 \,.
\end{equation}
Now, by applying the timelike geodesic condition $g_{ik}u^iu^k=-1$ into the
equation above, we find
\begin{equation} \label{eq:between}
f'\dot{v}\dot{r}=-\frac{1}{2}\varepsilon f' +\frac{1}{2}\varepsilon
ff'-\frac{1}{2} \varepsilon  f' \frac{L^2}{r^2} \dot{v}^2 \,.
\end{equation}
Substituting the equation~\eqref{eq:between} into
\eqref{eq:radialgeo} we arrive at the following general equation of
motion in terms of the metric function $f$ for the radial
coordinate
\begin{equation} \label{eq:geodesic}
\ddot{r}=   -\frac{1}{2}\left(1+\frac{L^2}{r^2} \right)
f'+f\frac{L^2}{r^3} \,.
\end{equation}
Hence, using the obtained metric function~\eqref{eq:methair}, one obtains the geodesic
equation in the form
 \begin{eqnarray}  \label{eq:geodesic2}
\ddot{r}&=&\left(-\frac{M}{r^2}+\frac{L^2}{r^3}-\frac{3ML^2}{r^4}
\right)_{sch}\nonumber\\
&&+ \left(-\gamma \frac{N}{2r^{\gamma+1}}-\left (\gamma+2
 \right)
\frac{NL^2}{2r^{\gamma+3}} \right)_s \nonumber \\
&&+\left(\frac{\alpha}{2M-\alpha l} e^{-\frac{2r}{2M-\alpha
l}}+ \frac{ \alpha L^2}{\left(2M-\alpha l\right) r^2}
e^{-\frac{2r}{2M-\alpha
l}}-\frac{\alpha L^2}{r^3} e^{-\frac{2r}{2M-\alpha l}} \right)_{h} \,,
\end{eqnarray}
where $\gamma=3\omega+1$.
From~\eqref{eq:geodesic2}, one can observe the following interesting points.
\begin{enumerate}
\item The three terms in the first line are the same as that
of the standard Schwarzschild black hole in which the first term
represents the Newtonian gravitational force, the second term represents
the repulsive centrifugal force, and the third term is the relativistic
correction of Einstein's general relativity, which accounts for the
perihelion precession.
\item The terms in the second line are new correction terms due to
the presence of the background field, which surrounds the hairy
Schwarzschild black hole, in which its first term is similar to the term
of the gravitational potential in the first brackets, while its second term
is similar to the relativistic correction of general relativity.
Then,
regarding~\eqref{eq:geodesic2} one realizes that for the more realistic
nonempty backgrounds, the geodesic equation of any object depends
strictly not only on the mass of the central object of the system and
the conserved angular momentum of the orbiting body,  but also on the
background field nature.
The new correction terms may be small in
general, in comparison to their Schwarzschild counterparts (the first and
the third term in the first brackets).
However, one can show that there
are possibilities that these terms are comparable to them.
One also can observe, by using the equation \eqref{eq:den}, that for $\omega
\in (-\frac{1}{3},\, 0)$ the Newtonian gravitational force is
strengthened by corrections caused by the surrounding field, on the
other hand, for other values of $\omega$ the force is weakened.
If we
consider the same question regarding the second term, which corresponds
to the relativistic correction of Einstein's general relativity, then
for values $\omega\in(-1,\, 0)$ the force is strengthened and this is while
  this force is weakened for other values $\omega$.
The
surrounding fluid does not have any contributions to the
repulsive centrifugal force.
\item The terms in the third line represent modifications  by
the primary hairs $\alpha$ and $l$.
The second term here corresponds to the
relativistic correction of Einstein's general relativity.
The third term here
represents a new correction  by the primary hairs to the
repulsive centrifugal force.
One can define the effective distance $D$
to find out  where this force disappears by relation
$\frac{A_1}{A_r}\approx 1$  where $A_r$ is the Schwarzschild black hole
repulsive centrifugal force, and $A_1$ is the correction to this force
caused by primary hairs.
So the distance is given by
\begin{equation} \label{eq:distance}
 D=\left (M-\frac{\alpha l}{2} \right )\ln \alpha \,.
\end{equation}
Considering a minimal geometrical deformations, $\alpha$ must be
negligible, i.e.,\, $\alpha \ll 1$.
So according to \eqref{eq:distance}, the correction caused by
primary hairs can weaken
the repulsive centrifugal force but it cannot cancel it, and hence this
correction is negligible, in general.
The first term in \eqref{eq:geodesic2} contributes a correction
to the Newtonian potential. This  can be seen using the effective potential
$V_{eff}(r)$.  One can write the geodesic equations in the form
\begin{equation} \label{eq:effective_potential}
V_{eff}(r)=\Phi(r)+\frac{L^2}{2r^2}+\Phi(r)\frac{L^2}{r^2} \,,
\end{equation}
where $\Phi(r)$ is related to $g_{00}$ metric component via relation
\begin{equation}
g_{00}=-\left(1+2\Phi \right) \,.
\end{equation}
By comparing this with \eqref{eq:methair}, we come to the conclusion that
\begin{equation}
\Phi(r)=-\frac{M}{r}+\frac{N}{2r^{3\omega+1}}-\alpha e^{-\frac{r}{2M-\alpha l}} \,.
\end{equation}
Now, taking the derivative of $V_{eff}$ in \eqref{eq:effective_potential}
with respect to $r$ \begin{equation}
\frac{d^2r}{d\tau^2}=-\frac{dV_{eff}}{dr} \,,
\end{equation}
we arrive at the equation  of motion \eqref{eq:geodesic2}.
\end{enumerate}

In order to better understand the  nature of the solution obtained in \eqref{eq:methair},
one can consider the following two groups of forces and
investigate their behaviour for various set of surrounding fields and
primary hair parameters.  
\begin{equation} \label{eq:g}
G\equiv \frac{M}{r^2}+\gamma
\frac{N}{2r^{\gamma+1}}-\frac{\alpha}{2M-\alpha l}
e^{-\frac{2r}{2M-\alpha l}} \\,
\end{equation}
\begin{equation} \label{eq:h}
H\equiv \frac{3ML^2}{r^4}+\left (\gamma+2 \right )
\frac{NL^2}{2r^{\gamma+3}}-\frac{\alpha L^2}{(2M-\alpha l) r^2}
e^{-\frac{2r}{2M-\alpha l}} \,,
\end{equation}
 where  $G$ group represents the Newtonian
gravitational force with its modifications and   $H$ group  corresponds to the
relativistic corrections of the general relativity.
One can ask for the possibilities  if the new modifications caused by surrounding fields and
primary hairs can cancel the original forces or change their effect, i.e
., change
their sign.
Hence, we are interested  in possible  cases in which for set of parameters $\omega$,
$\alpha$ and $l$, the $G$ and $H$ functions are getting negligible values
or they change their signs.
In the following subsections, we consider some specific fields possessing particular equations
of state motivated by cosmology.

However, we can note the following facts which we can derive from
\eqref{eq:h}. Let us consider the first two terms: for $-1<\omega<0$
these two terms are always positive. However, the second term is
negative for positive $\omega$, and we can expect the sign change of $H$.
Let us consider two particular cases:
\begin{itemize}
\item The radiation $\omega=\frac{1}{3}$. In this case, $|N|\leq M^2$
and the first two terms become negative in the region $0\leq r\leq 2M/3$,
which is inside the event horizon. Because the third term in
\eqref{eq:h} is negligible we can conclude that $H$ is always positive
outside the event horizon region.
\item The stiff fluid $\omega=1$. In this case we can put $N=-M^4$ then
$f(r=M)>0$. Thus, in this case the event horizon location at the radius
is less than $M$. However, the first two terms in \eqref{eq:h}
become negative at $r=M$ and $H<0$ outside the event horizon region.
\end{itemize}

%%%%%%%%%%%%%%%%%%%%%%%%%%%%%%%%%%%%%%%%%%%%%%%%%%%%%%%%%%
\subsection{Stiff Fluid}

We begin our analysis of timelike geodesics with the surrounding fluid having
 the  average equation of
the state
of a stiff fluid as\begin{equation}
P=\rho\, \Leftrightarrow\, \omega =1.
\end{equation}
As mentioned previously, the presence of the surrounding field has a weakening effect on the forces given by~\eqref{eq:g} and~\eqref{eq:h}.
From~\eqref{eq:den}, one observes that $N$ must be negative to maintain a positive energy density for the surrounding fluid.
Our objective is to determine whether the corrections  by the surrounding field and primary
hairs can cancel out the initial Schwarzschild forces or potentially can
change their sign, and thereby, altering the direction of the forces.

\begin{figure}[h!]
    \centering
 \includegraphics[scale=0.39]{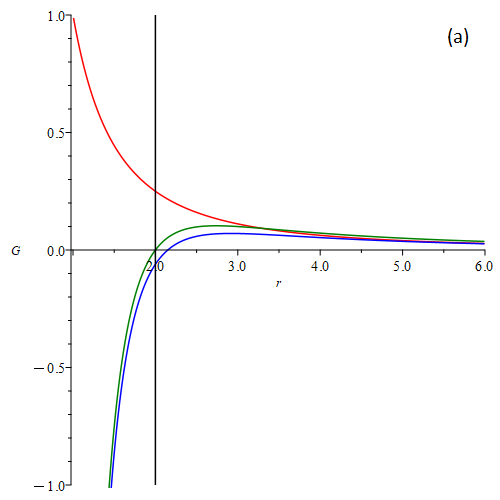}
 \includegraphics[scale=0.39]{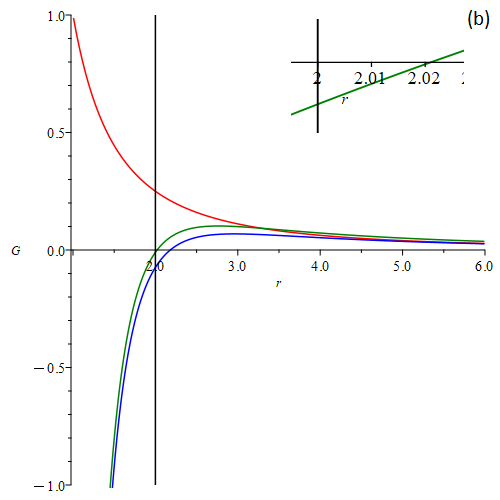}
 \includegraphics[scale=0.39]{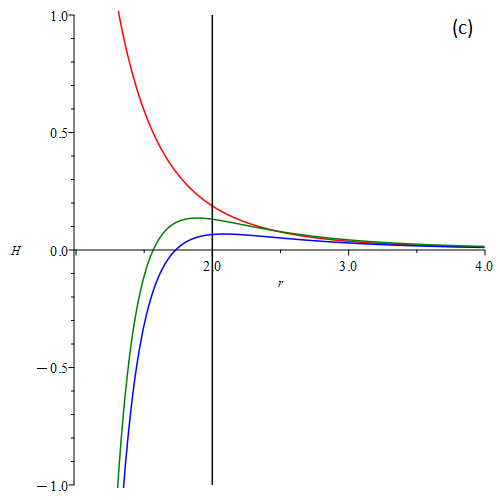}
    \caption[fig1]{Plot (a) shows the function $G$ versus the distance
        $r$ for $N=-4.972,\, l=1.514,\,
\alpha=0.5$ and $\mathcal{M}=1$.  Plot(b) shows the function $G$ versus the distance $r$
    for  $N=-5.186,\, l=1.567,
\alpha=0.5$ and $\mathcal{M}=1$, the small picture shows the function
        $G$ of hairy Kiselev black hole in the horizon vicinity. Plot
        (c)
        shows the
        function $H$ versus
        the distance $r$ for  $N=-5.186,\, l=1.567,
\alpha=0.5$ and $\mathcal{M}=1$. The red, blue, and green curves represent
 the
    Schwarzschild,  Kiselev, and 
    hairy Kiselev cases, respectively.}
    \label{fig:1}
\end{figure}

In Figure \ref{fig:1}(a), we plotted three curves corresponding the usual Schwarzschild,
Kiselev and hairy Kiselev black holes.
We observe that the function $G$
for the hairy Kiselev black hole is negligible but positive near the event
horizon $r=2\mathcal{M}$ for the given specific set of parameters.
However, in the case of purely Kiselev
black hole (i.e., $\alpha=0$), the function $G$ is negative in the
interval $2\leq r \leq 2.15$. One notes that in the purely Kiselev case, we have
a naked singularity (NS) (i.e., $g_{00}\neq 0$).\footnote{For this set of
parameters $g_{00}$ is always negative, i.e., there are not positive roots
of the equation $g_{00}=0$  for $r\in(0,+\infty)$. On
this reason, we have concluded that $r=0$ represents a NS because the
Kretschmann scalar  diverges at $r=0$.
By NS we mean that $r=0$ singularity is not covered with
the event horizon. The question about future-directed non-space-like
geodesics, which terminate at this singularity in the past, has not been
considered within this paper.}

Figure \ref{fig:1}(b) shows that the function $G$ becomes negative in
the vicinity of the event horizon (i.e.,\ in the region $2\leq r \leq 2
.02$) for the hairy
Kiselev black hole for the set of parameters $N = -5.186,\, l = 1.567$.
To have a bigger distance from the event horizon, where the function $G$ can become negative, one should increase $|N|$ and $l$, however, in this
case, $\mathcal{M} \sim \alpha l/2$ and it will not anymore be a minimal
geometrical deformation in~\eqref{eq:hairysch}.
So we can conclude that
$G$ might be negative outside the event horizon but only in its vicinity.

Figure \ref{fig:1}(c) compares the function $H$ for the Schwarzschild, Kiselev
and hairy
Kiselev cases for the values considered in Figure 1(b).

\begin{figure}[h!]
 \centering
 \includegraphics[scale=0.28]{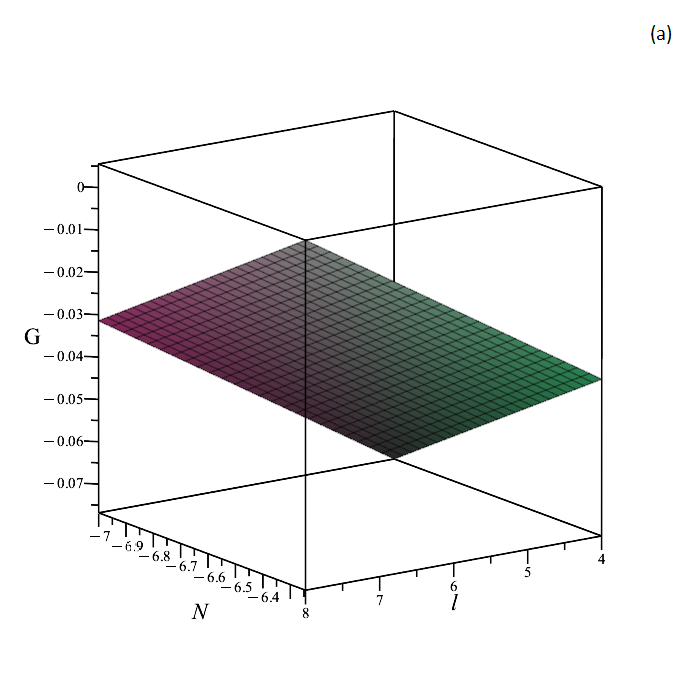}
\includegraphics[scale=0.28]{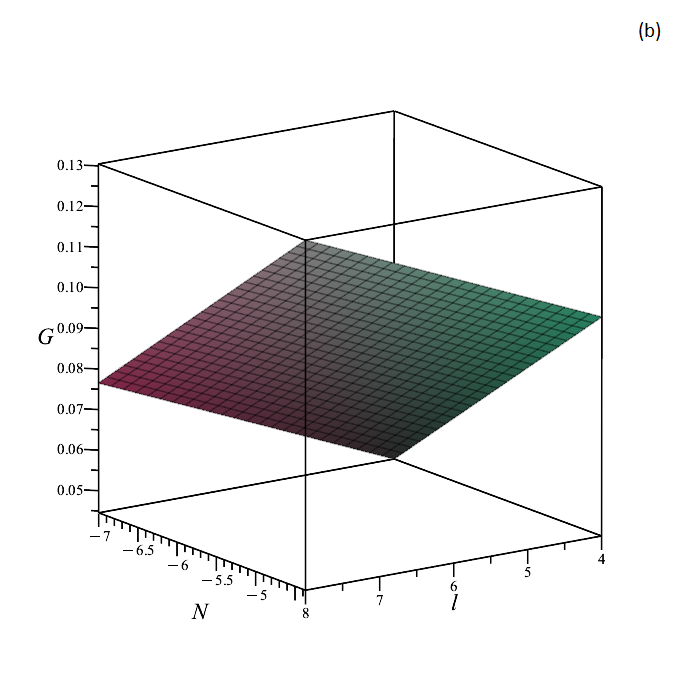}
 \includegraphics[scale=0.28]{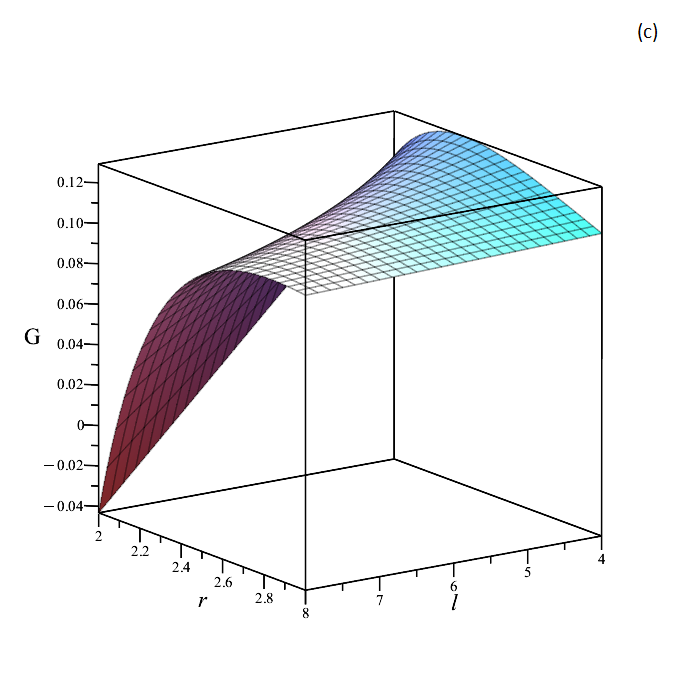}
\caption[fig2.2]{Plot(a) shows the function $G$ versus the parameters $N\in [-7, -6.245],\, l\in [4,8]$ for $r=2.1,\,
\alpha=0.1$ and $\mathcal{M}=1$. Plot(b) shows the function $G$ versus the parameters $N\in[-7, -4.367],\, l\in [4,8]$ for $r=2.5,\,
\alpha=0.1$ and $\mathcal{M}=1$. Plot(c) shows the function $G$ versus $l\in [4,8],\, r\in[2,3]$
    for  $N\in [-6.183,-2.983],\, \alpha=0.1$ and $\mathcal{M}=1$. The event horizon,  located
        at $r=2\mathcal{M}$, follows the condition $N=- 0
        .8 l+ 1.6 e^{-2}$.}
    \label{fig:2}
\end{figure}

In order to understand better the influence of a primary hair on a
geodesic motion we put $\alpha=0.1$ in order to consider bigger values
of $l$.
Figures \ref{fig:2}(a) and \ref{fig:2}(b) show how $G$ changes with
different values of $l$ and $N$.
One can see that there are regions where it
becomes negative.
However, from these pictures one cannot realize if they
deal with a black hole or a naked singularity.
For this purpose one should
impose the condition of existence of an event horizon .
The Figure \ref{fig:2}(c) shows how $G$ changes in this case.

%%%%%%%%%%%%%%%%%%%%%%%%%%%%%%%%%%%%%%%%
\subsection{Radiation}

Here we consider the surrounding field having the average equation
of state of radiation field as
\begin{equation}
P=\frac{\rho}{3}\, \Leftrightarrow\, \omega =\frac{1}{3} \,.
\end{equation}
In this case, the $N$ parameter must be negative, and akin to the previous
case, the  surrounding radiation field and primary hairs weaken the forces~in
\eqref{eq:g}
and~\eqref{eq:h}.

\begin{figure}[h!]
    \centering
    \includegraphics[scale=0.39]{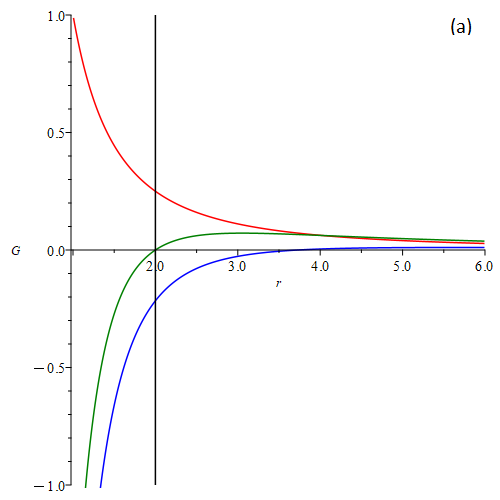}
        \includegraphics[scale=0.39]{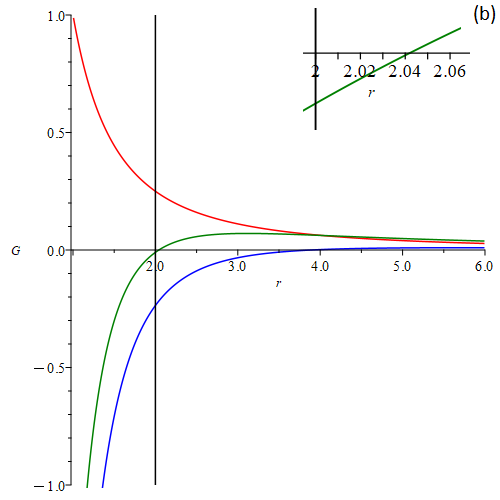}
            \includegraphics[scale=0.39]{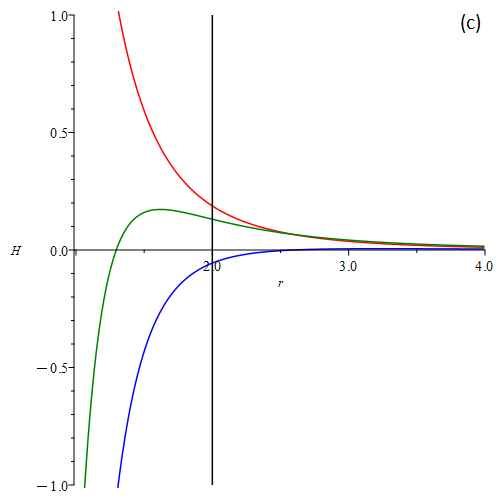}
    \caption[fig4]{Plot(a) shows the function $G$ versus the distance $r$    for  $N=-3.729,\, l=4,\,
\alpha=0.5$ and $\mathcal{M}=1$. Plot(b) shows  $G$ versus $r$
    for $N=-3.889,\, l=4.16,\,
\alpha=0.5$ and $\mathcal{M}=1$, the small picture shows the function
        $G$ of hairy Kiselev black hole in the horizon vicinity. Plot(c)
        shows  $H$ versus $r$
    for  $N=-3.889,\, l=4.16,\,
\alpha=0.5$ and $\mathcal{M}=1$.  The red, blue, and green curves
        correspond to the
    Schwarzschild, Kiselev, and hairy Kiselev cases, respectively.}
    \label{fig:3}
\end{figure}

Figure \ref{fig:3}(a) shows three curves in the pure Schwarzschild, Kiselev
and hairy Kiselev black holes for  the parameter values $N=-3.729$ and $l=4$.
For the case of surrounding radiationlike field, one observes that the
spacetime is akin to the hairy Reissner-Nordstrom black hole such that
the parameter $N$ plays the role of black hole's electric charge, i.e.
$N=-Q^2$.
So, in purely Reissner-Nordstrom case, the curve corresponds to the
naked singularity because $\mathcal{M}^2<Q^2$.
In comparison to the stiff fluid case, one notes
that the parameters $l$ and $N$
are taken greater values to ensure that the function $G$ is negligible.

In Figure \ref{fig:3}(b), we plotted curves in order to show that hairs
can affect
the geodesic motion and hence $G$ can become negative in the event horizon
vicinity (in the region $2\leq r \leq 2.042$).
In this case, we set $N=-3.889$
and $l=4.16$. 
One can see that the smaller values of $\omega$ we take, the bigger values
of $l$ are required to ensure the negative values of $G$.
For example, if we
take this value of $l$ (i.e., $l=4.16$), then in the case of stiff
fluid, we have $N=-15.557$ (we obtain this value by demanding that
the event horizon is located at $r=\mathcal{M}$), then the $G$ function
is negative in the region $2\leq r \leq 2.534$.
Thus, one can see that
the region, where negative values of $G$ are possible,  shrinks when
$\omega$ tends to zero.
 Figure \ref{fig:3}(c) denotes the function $H$ with the values of $N$
and $l$ as in  the previous figure.

\begin{figure}[h!]
    \centering
    \includegraphics[scale=0.28]{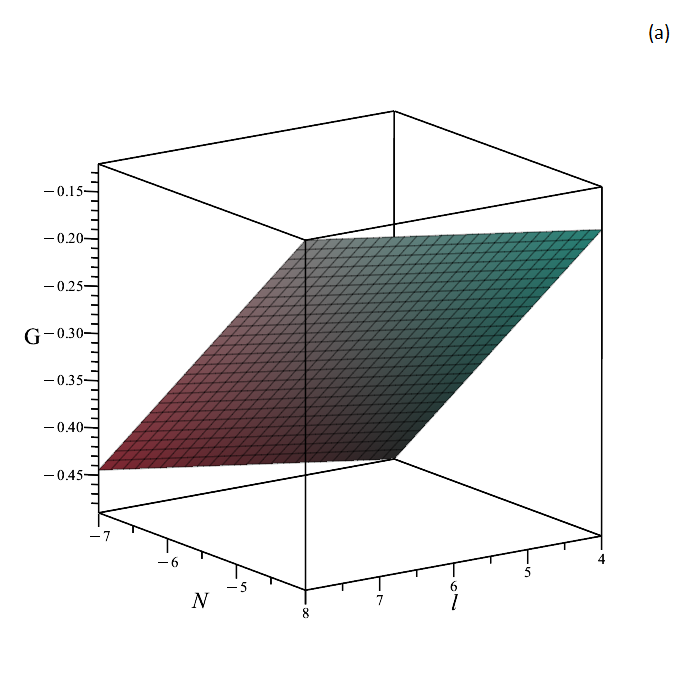}
        \includegraphics[scale=0.28]{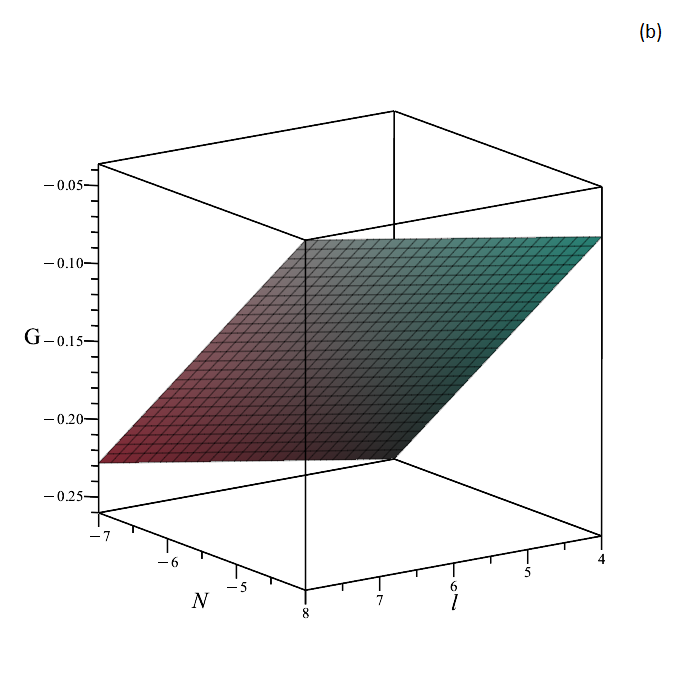}
       \includegraphics[scale=0.28]{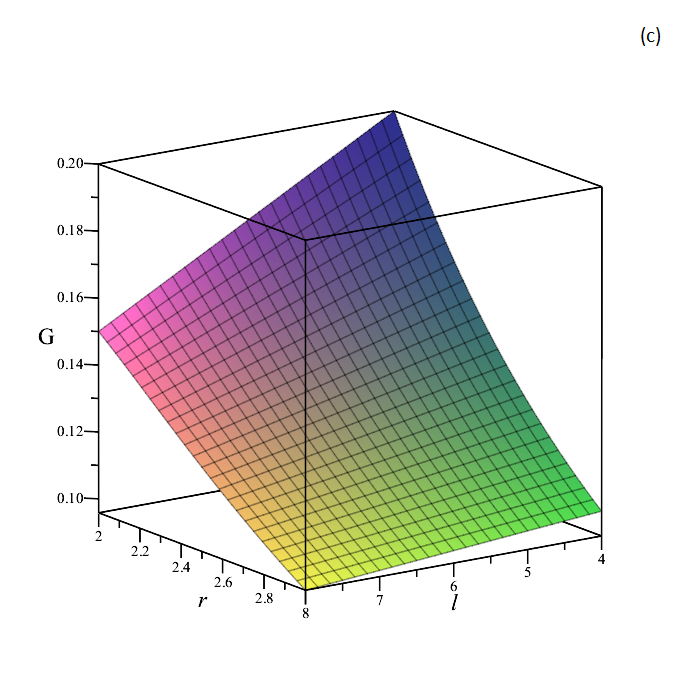}
    \caption[fig2.2]{Plot(a) shows the function $G$ versus the  parameters
  $N\in[-7,-4],\, l\in [4,8]$ for $r=2.1,\,
\alpha=0.1$ and $\mathcal{M}=1$. Plot(b) shows the function $G$ versus the  parameters  $N\in[-7,-4],\, l\in[4,8]$ for $r=2.5,\,
\alpha=0.1$ and $\mathcal{M}=1$.
Plot(c) shows  the function $G$ versus  $r,\,l$
    for  $N\in [-1.546,-0.746],\, \alpha=0.1$ and $\mathcal{M}=1$.  The event horizon,  located
        at $r=2\mathcal{M}$, must satisfy the condition $N=- 0.2 l+ 0.4 e^{-2}$.}
    \label{fig:4}
\end{figure}

Similar to  the stiff fluid  case, we have several plots for $\alpha=0.1$.
Figures \ref{fig:4}(a) and \ref{fig:4}(b) show that $G$
becomes
negative at the larger distances in comparison to the stiff fluid case. This
apparently contradicts our previous statement that the smaller $\omega$ we consider,  the region where $G$ becomes negative becomes smaller.
However, one notes that this is a case
of the naked singularity because if one imposes an
extra condition of the event horizon existence, then for this case
($\alpha=0.1$) the $G$ function is always positive outside the horizon
as can be seen from Figure \ref{fig:4}(c).

%%%%%%%%%%%%%%%%%%%%%%%%%%%%%%%%%%%%%%%%%%
\subsection{Dust}

For a  dustlike field we have \begin{equation}
P=0 \, \Leftrightarrow\, \omega=0 \,,
\end{equation}
and we can show analytically that the function $G$ is positive near the
event horizon as follows.
We have
\begin{equation}
\frac{2M+N}{r}=1+\alpha e^{-\frac{r}{\mathcal{M}}} \,.
\end{equation}
Substituting this into \eqref{eq:g} and considering the event horizon at
$r=2\mathcal{M}$, one obtains
\begin{equation}
\frac{1}{4\mathcal{M}}-\frac{\alpha}{4\mathcal{M}e^2}>0 \,.
\end{equation}
So, for physically relevant values of $\alpha,\, l$ and $N$, the
function $G$ is positive outside the event horizon.

\begin{figure}[H]
    \centering
    \includegraphics[scale=0.35]{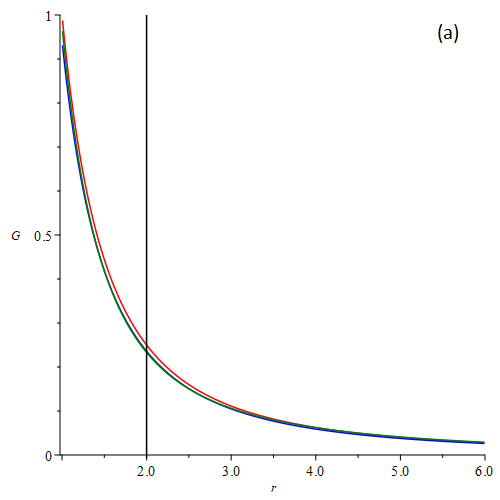}
        \includegraphics[scale=0.35]{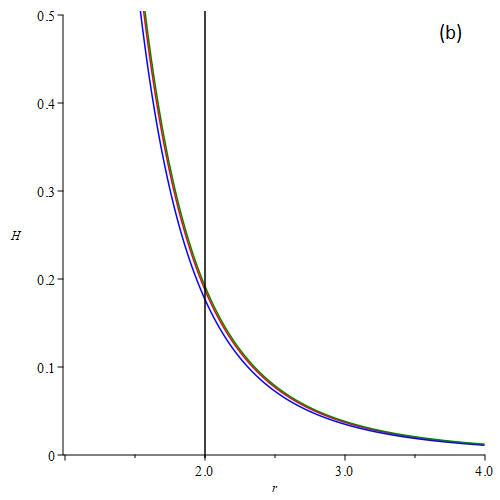}
              \includegraphics[scale=0.32]{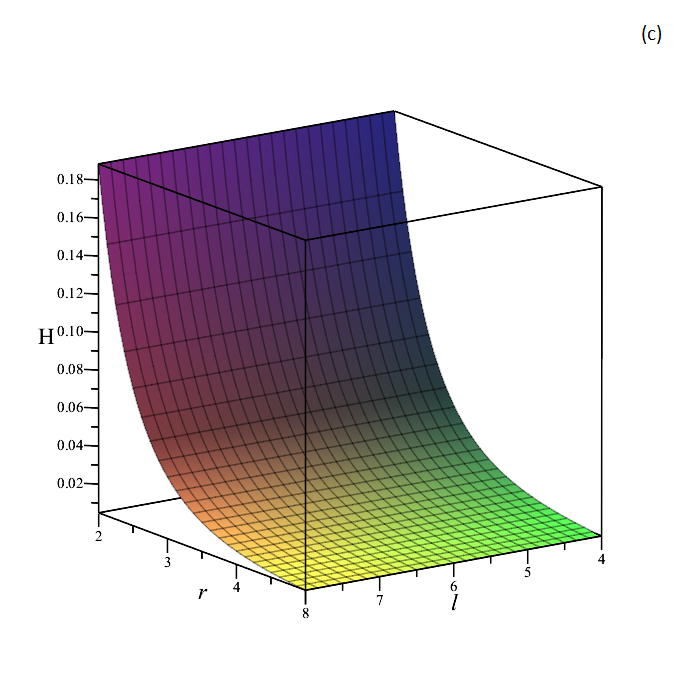}
    \caption[fig7]{Plot(a) shows the function $G$ versus the distance $r$ for $N=-0.115,\, l=0.5,
\alpha=0.5$ and $\mathcal{M}=1$. Plot(b) shows the function $H$ versus the distance $r$
    for  $N=-0.115,\, l=0.5,\,
\alpha=0.5$ and $\mathcal{M}=1$. The red, blue, and green curves
        correspond to the  Schwarzschild, Kiselev, and hairy Kiselev cases, respectively. Plot(c) shows the function $H$ versus  $r\,,l$
    for the values $N\in [-0.773,-0.373],\, l\in[4,8],\, r\in[2,5],\,
\alpha=0.1$ and $\mathcal{M}=1$.  The event horizon,  located
        at $r=2\mathcal{M}$, must satisfy the condition $N=- 0
        .1 l+ 0.2 e^{-2} $. }
    \label{fig5}
\end{figure}

Figure \ref{fig5}(a) compares  three curves of a hairy
Kiselev black hole, purely Kiselev when $\alpha=0$, and Schwarzschild
case when $\alpha=0$ and $N=0$.
These curves are plotted for  $l=0.5\,, N=-0.115$.
Figure \ref{fig5}(b)
is plotted for the same values of black hole parameters and shows the
behaviour of the function $H$.
For $\omega\geq 0$ the function $H$ is positive, and  its
behaviour is shown in the Figure \ref{fig5}(c).
For other values of $\omega$ we
could not find the condition (at small values of $\alpha$) where $H$
becomes negative.

%%%%%%%%%%%%%%%%%%%%%%%%%%%%
\subsection{Quintessence}

\begin{figure}[h]
    \centering
    \includegraphics[scale=0.43]{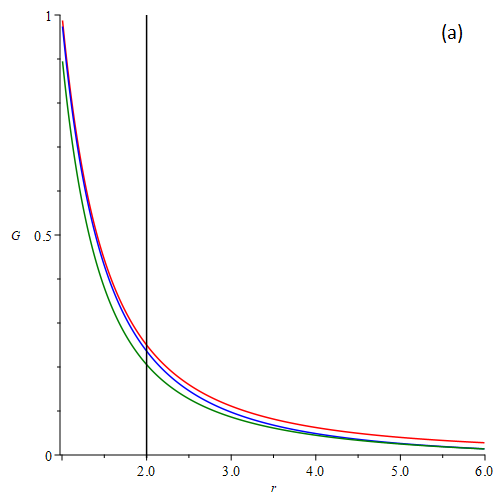}
        \includegraphics[scale=0.43]{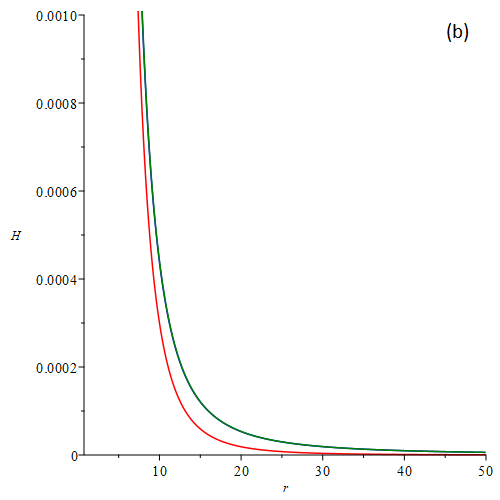}
    \caption[fig9]{Plot(a) shows the function $G$ versus the distance $r$
    for $N=0.028,\, l=0.05,
\alpha=0.5$ and $\mathcal{M}=1$. Plot(b) shows the function $H$ versus
        the distance $r$  for $N=0.028,\, l=0.05,\,\alpha=0.5$ and $\mathcal{M}=1$. The red, blue, and green curves correspond to the  Schwarzschild, Kiselev, and hairy Kiselev cases, respectively.}
    \label{fig:9}
\end{figure}

For a quintessencelike field, the equation of the state is
\begin{equation}
P=-\frac{2}{3}\rho \, \Leftrightarrow\, \omega=-\frac{2}{3} \,.
\end{equation}
In this case, the parameter $N$ must be positive as one can see from
\eqref{eq:den}.
The function $G$ can be negligible in the vicinity of
the horizon only if either $N$ or $L$ are negative.
However, $G$ can
take negative values but at large distances from the event horizon. As
can be shown from Figure \ref{fig:9}(a) at values $l=0.05\,, N=0.028$, the
function
$G$ for Kiselev black hole becomes negative at $r>8.553$. The effect of
$N$ and $\alpha$ on the function $H$ for these values are negligible and
they become considerable only at large distances, as one can see from Figure
\ref{fig:9}(b) .

%%%%%%%%%%%%%%%%%%%%%%%%%%%%%%%%%%%%%%%
\subsection{De Sitter background}

In this case, the surrounded  fluid has the effective equation of the state
\begin{equation}
P=-\rho \, \Leftrightarrow\, \omega=-1 \,.
\end{equation}
Like in the previous case, the parameter $N$ must be negative, and the
function $G$ must be positive near the event horizon.

\begin{figure}[H]
    \centering
    \includegraphics[scale=0.39]{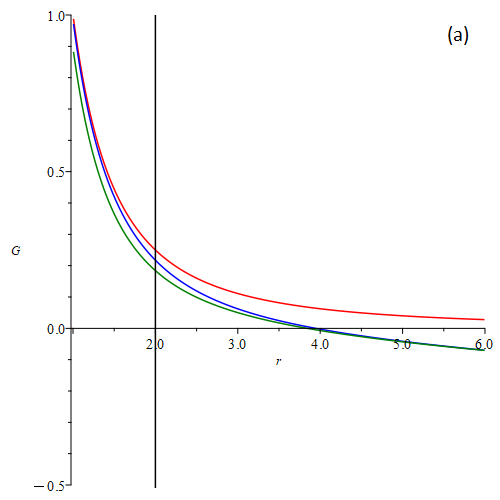}
        \includegraphics[scale=0.39]{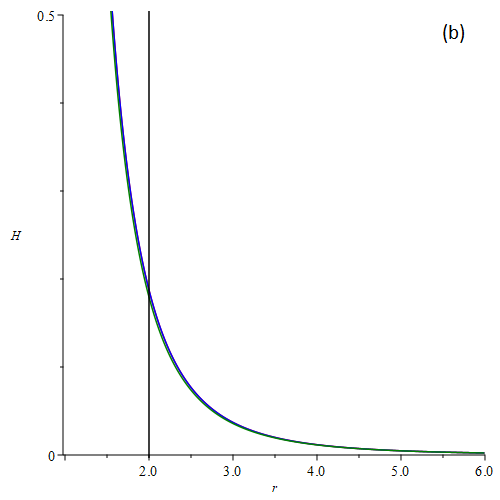}
    \caption[fig11]{Plot(a) shows the function $G$ versus the distance $r$
    for $N=0.016,\, l=0.01,\,\alpha=0.5$ and $\mathcal{M}=1$.Plot(b) shows the function $H$ versus the distance $r$ for the same values of parameters.
The red, blue, and green curves correspond to the  Schwarzschild,
        Kiselev, and hairy Kiselev cases, respectively. }
    \label{fig:11}
\end{figure}

Figure \ref{fig:11}(a) shows that the function $G$ for $N=0.016\,, l=0.01$
becomes negative for $r>3.841$.
The function $H$ behaves very similar in all three cases as can be seen in Figure \ref{fig:11}(b).

\begin{figure}[H]
    \centering
    \includegraphics[scale=0.3]{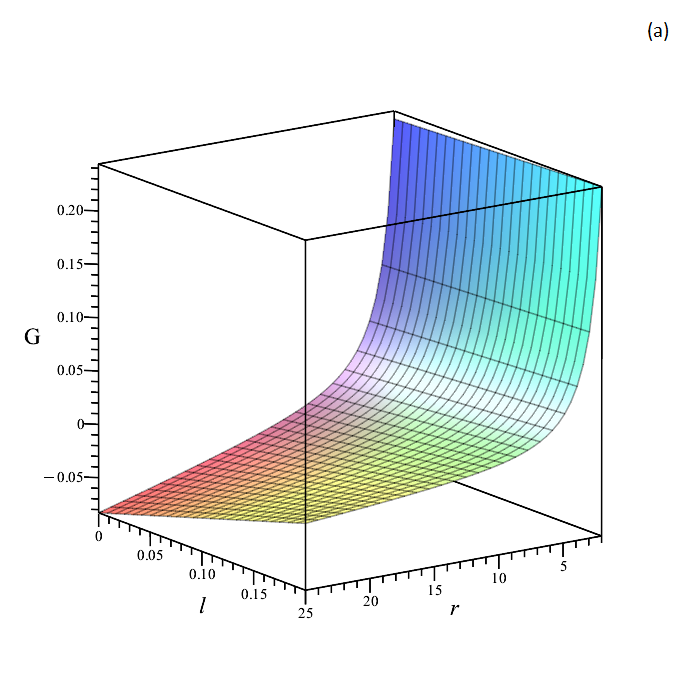}
        \includegraphics[scale=0.3]{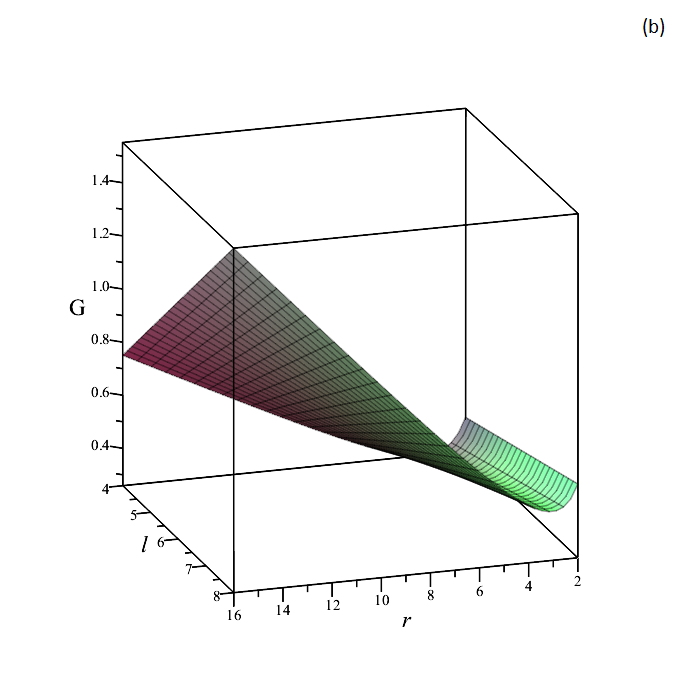}
    \caption[fig2.2]{Plot (a): the dependence of the function $G$ on the
        $r\,,l$
    for the values $N=0.003..0.0008 \,, l=0..0.2 \,, r=2..25\,,
\alpha=0.1$ and $\mathcal{M}=1$. The event horizon, when it is located
        at $r=2\mathcal{M}$, must satisfy the following condition $N=- 0
        .0125 l+ 0.025 e^{-2} $.  Plot (b): the dependence of the
        function $G$ on the $r\,,l$
    for the values $N=-0.047..-0.097 \,, l=4..8 \,, r=2..16\,,
\alpha=0.1$ and $\mathcal{M}=1$. The event horizon, when it is located
        at $r=2\mathcal{M}$, must satisfy the following condition $N=- 0
        .0125 l+ 0.025 e^{-2} $.}
    \label{fig:2.8}
\end{figure}
Figure \ref{fig:2.8} shows the behaviour
of $G$ at
$\alpha=0
.1$ and with an extra condition of the event horizon existence.
Here
\ref{fig:2.8}(a) is plotted for positive cosmological constant as
\ref{fig:2.8}(b)
for negative cosmological constant - anti-de-Sitter case.

\subsection{Phantom field}

\begin{figure}[H]
       \centering
    \includegraphics[scale=0.39]{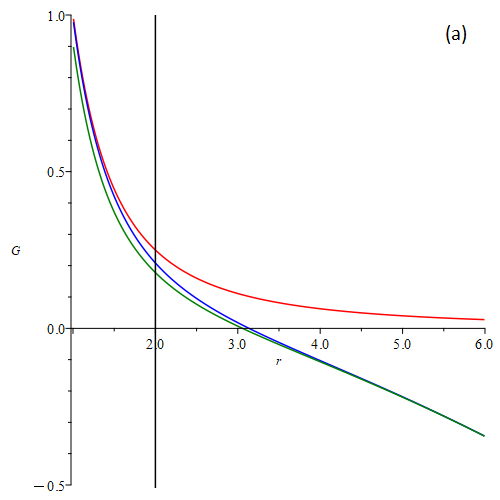}
        \includegraphics[scale=0.39]{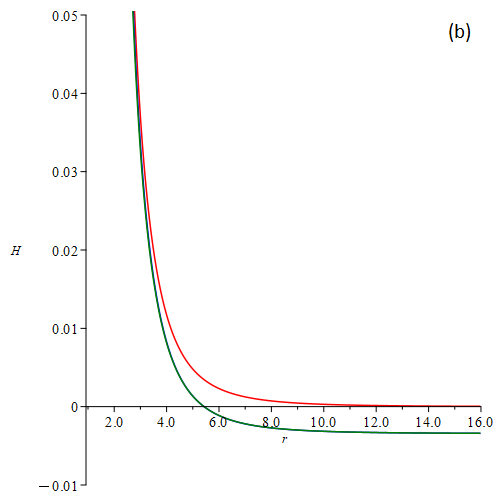}
    \caption[fig13]{Plot(a) shows the function $G$ versus the distance $r$
    for $N=0.007\,, l=0.05,
\alpha=0.5$ and $\mathcal{M}=1$. Plot(b) shows the function $H$ versus the distance $r$
    for $N=0.007\,, l=0.05,
\alpha=0.5$ and $\mathcal{M}=1$. The red, blue, and green curves
        correspond to the  Schwarzschild, Kiselev, and hairy Kiselev cases, respectively.  }
    \label{fig:13}
\end{figure}

In general, the equation of the state a phantomlike field lies in the range $\omega<-1$
\cite{bib:cosd_phf,bib:phf_bs,bib:tach_phf, ph1, ph2, ph3, ph4}. In order to study the effect
of a phantom field, one can consider, as instance
\begin{equation}
P=-\frac{4}{3}\rho \, \Leftrightarrow\, \omega = -\frac{4}{3}\,.
\end{equation}
The parameter $N$ must be positive, and as can be seen in Figure
\ref{fig:13}(a), the
function $G$ takes negative values at the region $r>3.056$ at $l=0.05\,,
N=0.007$.
Figure
\ref{fig:13}(b) shows that for the same values of $l$ and $N$, the
function $H$ can be negative in the region $r>5.433$.

\section{Conclusion}

Inspired by the fact that black holes inhabit nonvacuum  cosmological
backgrounds, we present a new solution to the Einstein field equations   representing a surrounded
hairy
Schwarzschild black hole.
This solution
takes into account both the primary hair and surrounding  fields (represented
by an energy-momentum tensor following the linearity and additivity condition
\cite{bib:50}), which affect the properties of the black hole.
The effect of the corresponding contributions
on timelike geodesics are discussed. We find that the new induced modifications
can be considerable in certain cases.
In particular, we investigate how
the specified surrounding fields and primary hairs affect the Newtonian and
perihelion precession terms. Our observations are as follows.

\begin{itemize}
\item The surrounding fields with 
$-\frac{1}{3}<\omega<0$ contribute positively to the Newtonian term, i.e
., strengthening the gravitational attraction.
\item The new corrections to the Newtonian term might be the same order or
even greater for all other cases  if one considers a naked singularity
.\footnote{Considering the positive $\omega$, the weak energy condition
demands negative  $N$ values. This restriction, for example, in
the dust case  requires $|N|<2M$, otherwise, the metric function $f(r)$ is
always positive for all
ranges of $r$ since all the being four terms are positive, and hence there is no event
horizon. In the case of the radiation, i.e., $\omega=\frac{1}{3}$, the
NS occurs if $M^2+N<0$ which requires large values of
    $|N|$. Hence one observes that for bigger values of $|N|$, the
    function $|G|$ becomes bigger, but
this implies the violation of the condition required for the existence of an event horizon.}
\item In the case that the solution represents a  black holes, new corrections  can be of the same order or
even greater than the Newtonian term in the event
horizon vicinity for $\omega>0$.
\item For $\omega<-\frac{1}{3}$, i.e., for effectively repulsive fluids
akin
to dark energy models, the correction terms dominate far from the event horizon and mainly near the
cosmological horizon.
\end{itemize}

The Schwarzschild black hole is an idealized vacuum solution, and it is
important to consider how it gets deformed in the presence of matter fields.
Another
crucial factor to consider is the impact of the surrounding environment,
particularly the shadow of a black hole in the cosmological background,
which serves as a potential cosmological ruler~\cite{bib:ruller}.
The
solution presented in this work can be further investigated to study the
shadow of a hairy Schwarzschild black hole in various
cosmological backgrounds in order to find out how anisotropic
fluid can affect the observational properties~\cite{bib:hor_sc_eht},
which
is a
plan of
our
upcoming investigations.
It is worthwhile to mention that  applying the Newman-Janis~\cite{bib:71r} and
Azreg Ainou~\cite{bib:73r, bib:74r} algorithms one can obtain the
rotating version of the solution presented here. Also, investigation of quasinormal modes, thermodynamics properties,
accretion process, and gravitational lensing of these solutions can help
us to understand better the nature of these objects.

 The obtained hairy Kiselev solution has many potential uses in various
cosmological and astrophysical scenarios. It can be an arena for
high-energy phenomena. If one considers the centre of mass energy $E_{c
.m.}$ of two colliding particles in usual Schwarzschild spacetime, than the
value is quite limited and small~\cite{bib:bsw}. However, two extra
terms here might lead to the existence of the innermost stable equilibrium
point in the horizon vicinity~\cite{bib:zaslav_anti}, which can lead to
unbound centre-of-mass energy $E_{c.m.}$ of two colliding particles.
Another tool to distinguish the hairy Kiselev black hole from the usual
Schwarzschild one is to study its shadow properties. The shape of the
shadow is the same as in the Schwarzschild case due to the spherical
symmetry. However, the existence of four extra parameters $\omega\,, N\,
, \alpha\,, l$ have, surely, impact on its size and
intensity~\cite{bib:joshi_shadow}. The study of the planet's motion is
the way to define if a primary hair can have an impact on its trajectory
. As we have shown, extra terms can drastically change particle motion.
However, in a realistic astrophysical situation, one should consider
this motion near the black hole where $\alpha\,, l\,, N$ has a large
impact on the particle motion. Based on the parameters and variables
considered in this model, it seems that attempting to test it within the
Solar system would be futile. This is because any additional terms are
essentially insignificant beyond the surface of the Sun, resulting in a
prediction that would be indistinguishable from that which is already
predicted in Schwarzschild spacetime. Therefore, it may be more
beneficial to focus on the study of black hole vicinity where more
noticeable results can be achieved.
Although it has not yet been observed, the Hawking temperature and
radiation may get also influenced by a primary hair~\cite{bib:thermo,
    bib:ver_thermo}.
The Schwarzschild black hole possesses a negative heat
capacity. Cosmological fields and primary hair might lead to positive
specific heat capacity and phase transition~\cite{bib:kudr_bardeen}. All
these are the topics of our future investigations.
\\
\\
\textbf{Acknowledgments}: V. Vertogradov and M. Misyura say thanks to
grant NUM. 22-22-00112 RSF for financial support.

\end{document}